# Energy-looping and photon-avalanche-like phenomena in $Nd_xY_{1.00-x}Al_3(BO_3)_4$ powders excited at 1064 nm

Rodrigo F. da Silva,[a] Daniel F. Luz,[a] Célio V. T. Maciel,[a] Emanuel P. Santos,[b] Gabriela Soares,[a] Lauro J. Q. Maia,[c] Carlos Jacinto,[d] and André L. Moura[d,*]

[a]Grupo de Física da Matéria Condensada, Núcleo de Ciências Exatas – NCEx, Campus Arapiraca, Universidade Federal de Alagoas, Arapiraca-AL, Brazil

[b]Departamento de Física, Universidade Federal de Pernambuco, Recife-PE, Brazil

[c]Instituto de Física, Universidade Federal de Goiás, Goiânia-GO, Brazil

[d]Programa de Pós-graduação em Física, Instituto de Física, Universidade Federal de Alagoas, 57072-900, Maceió-AL, Brazil

*Corresponding author. E-mail: andre.moura@fis.ufal.br

**Abstract**

Recently, a photon-avalanche-like (PA-like) process based on trivalent neodymium ions ($Nd^{3+}$) under an unconventional excitation at 1064 nm was demonstrated using stoichiometric $NdAl_3(BO_3)_4$ particles. Although the $Nd^{3+}$ can emit radiation at 1064 nm efficiently, they present very small absorption at this wavelength due to the absence of resonant ground-state transitions. However, phonon-assisted excitation ($^4I_{9/2} \rightarrow {}^4F_{3/2}$ and/or $^4I_{11/2} \rightarrow {}^4F_{3/2}$) followed by cross-relaxation between one excited ion ($^4F_{3/2}$) with another at the ground state ($^4I_{9/2}$), and subsequent phonon emissions $\{^4I_{15/2}, {}^4I_{15/2} \rightarrow {}^4I_{13/2}, {}^4I_{13/2} \rightarrow {}^4I_{11/2}, {}^4I_{11/2}\}$, provide two ions at the $^4I_{11/2}$ state, from which can occur the resonant excited-state absorption to the $^4F_{3/2}$ level. Reestablishing the sequence of events, the absorption of photons at 1064 nm can increase greatly. Besides the emission around 880 nm ($^4F_{3/2} \rightarrow {}^4I_{9/2}$), there are thermal excitations to upper-lying states, with subsequent emissions in the visible and near-infrared regions (480 – 2000 nm). Here, we investigate the role of the $Nd^{3+}$ content on the PA characteristics in $Nd_xY_{1.00-x}Al_3(BO_3)_4$ particles with *x* ranging from 0.05 to 1.00. It is known that the replacement of $Y^{3+}$ by $Nd^{3+}$ into the $YAl_3(BO_3)_4$ crystalline structure can introduce strong modifications of the lattice properties as well as in the photoluminescence characteristics, such as luminescence concentration quenching and broadening of spectral lines. Despite that, we observe, for low *x* (≤ 0.20) an energy-looping preceding the PA-like that ensues for *x* ≥ 0.40. It is associated to the proximity among the $Nd^{3+}$ ions, fundamental to the electric dipole-electric dipole interaction responsible for the $Nd^{3+}$ energy transfer $\{^4F_{3/2}, {}^4I_{9/2}\} \rightarrow \{^4I_{15/2}, {}^4I_{15/2}\}$. We discuss the present results focusing on emerging technologies with development of ultra-sensitive thermal sensors, and super-resolution imaging thanks to the giant nonlinearities in the input-output power dependences.

1. INTRODUCTION

The replacement of trivalent Ytrium ions ($Y^{3+}$) by trivalent Neodymium ions ($Nd^{3+}$) in $YAl_3(BO_3)_4$ lattice, forming $Nd_xY_{1.00-x}Al_3(BO_3)_4$, is accompanied by changes in the structure and in the photoluminescence characteristics [1–4]. As well-described by Benayas et al. [1], for moderate $Nd^{3+}$ content ($x \leq 0.20$), the structural (hexagonal phase) properties are almost independent of $x$, as is the crystalline field around the $Nd^{3+}$ giving similar photoluminescence characteristics [1]. The lattice parameters and the photoluminescence features change significantly for $x$ in the range $0.20 < x < 0.75$. The increase in inhomogeneous broadening is attributed to changes in the crystal field by the inclusion of $Nd^{3+}$. A structural phase transition occurs for $0.75 < x \leq 0.90$. In this range, there are broadening of the photoluminescence spectra and discontinuities in the energy transfer parameters. The structural properties are disordered as in glasses, and the photoluminescence properties are almost independent of $x$. Finally, for $x > 0.90$ a monoclinic phase [$NdAl_3(BO_3)_4$] is established and sharp lines in the photoluminescence spectra unveil the high homogeneity of the crystal field.

Usually, as the $Nd^{3+}$ get closer by raising $x$, the probability of energy transfer between $Nd^{3+}$ pairs by cross-relaxations increases, which is usually deleterious for photoluminescence due to energy migration followed by energy lost to traps in the crystalline structure. This is called concentration quenching and has been observed in $Nd_xY_{1.00-x}Al_3(BO_3)_4$ single crystals and particle powders [1,3]. However, for $x > 0.90$, especially in the stoichiometric structure $NdAl_3(BO_3)_4$, the large quantum efficiency (0.3) of the $Nd^{3+}$ $^4F_{3/2}$ state is considered anomalous [5]. This implies a complex behavior from the structural and photoluminescence properties of $Nd_xY_{1.00-x}Al_3(BO_3)_4$ as $x$ is raised. In fact, laser and random laser emissions were observed in $Nd_xY_{1.00-x}Al_3(BO_3)_4$ single crystals and particle powders [4,6] with high efficiencies, and also the frequency conversions (self-second-harmonic, self-sum-frequency generation, and parametric amplification) [7–9], which is the laser/random laser generation followed by its conversions to other wavelengths in the interaction of the beams with the $Nd_xY_{1.00-x}Al_3(BO_3)_4$ crystalline structure. Interestingly, the stoichiometric structure $Nd_xY_{1.00-x}Al_3(BO_3)_4$ presents the best performance for both laser and random laser emissions. Under short-pulse excitation, there is an enhancement in the random laser performance when increasing $x$ [8], which is understood as the inhibition of the concentration quenching due to the formation of random laser pulses in short time scales ($\approx$10 ns) [10,11]. For laser and random laser emissions at around 1062 nm ($^4F_{3/2} \rightarrow {}^4I_{11/2}$), the $Nd^{3+}$ are usually excited at around 808 nm in resonance with the $^4I_{9/2} \rightarrow {}^4F_{5/2}$ transition. But, the excitation pathways can be with longer and shorter wavelengths. As an example, $NdAl_3(BO_3)_4$ shows the best performance when excited directly at the upper laser level ($^4F_{3/2}$) despite the lower absorbance when compared, for example, with the ground-state absorption (transition $^4I_{9/2} \rightarrow {}^4F_{5/2}$). This is due to the lower quantum defect between the excitation and lasing photon energies because the energy difference increases the particles temperature, which in turn decreases the laser performance due to the nonradiative relaxations from the $^4F_{3/2}$ state as well as thermal excitations to upper-lying levels. The $NdAl_3(BO_3)_4$ system presents two distinct crystalline fields for the $Nd^{3+}$, which can present laser emissions simultaneously at 1063.5 and 1065.1 nm [12]. Their distinct thermal responses allow them to operate as a ratiometric optical thermometer with high relative thermal sensitivity of 2.0 %°C$^{-1}$ in the temperature range from 20 °C to 70 °C [12]. In resume, as investigated in laser/random laser scenarios, $Nd_xY_{1.00-x}Al_3(BO_3)_4$ single crystals or particle powders exhibit complex characteristics that are interesting for photonics.

The unconventional excitation pathway of $Nd^{3+}$ at 1064 nm (intensities > 2×10$^4$ W/cm$^2$), presented by da Silva et al. [13], used, as a proof of concept, $NdAl_3(BO_3)_4$ particle powders. Although the $Nd^{3+}$ can emit radiation efficiently at around 1064 nm, due to the $^4F_{3/2} \rightarrow {}^4I_{9/2}$ transition, the absorbance at this wavelength is poor at room temperature, since this wavelength is nonresonant with ground-state transitions. However, by exploiting a photon-avalanche-like (PA-like) mechanism, the authors succeeded in demonstrating for the first time strong upconversion and downconversion from 480 nm to 2000 nm. Thus, the feasibility of operating an optical thermometer was demonstrated due to the large population redistribution among the $Nd^{3+}$ states. Since the excitation pathways involve creation and annihilation of phonons, in a subsequent work, Santos et al. [14] demonstrated the triggering of the PA-like when rising the particles temperature while keeping the excitation power fixed at a value below the threshold at room temperature.

First theoretically described [15], and recently demonstrated experimentally [16], the photon avalanche (PA) excitation of trivalent rare-earth ions can find applications in emerging technologies as imaging systems with resolution beyond the light diffraction thanks to the large nonlinearity of the input-output power dependence. The spatial resolution scales with $S^{-0.5}$, where $S$ is the number of photons annihilated to produce one emitted photon or, in other words, the degree of nonlinearity in the photoluminescence power ($P_{out}$) dependence with the excitation power ($P_{exc}$) – $P_{out} \propto P_{exc}^S$. Lee et al. [16] succeed in demonstrate the generation of images with spatial resolution of 70 nm using $Tm^{3+}$-doped $NaYF_4$ nanoparticles as a result of high nonlinearities ($S > 20$). This spatial resolution is far below the diffraction limit of the 1064 nm excitation beam as well as the 800 nm emission attributed to the $Tm^{3+}$ transition $^3H_4 \rightarrow {}^4H_6$.

Here, we investigate the PA-like characteristics of $Nd_xY_{1.00-x}Al_3(BO_3)_4$ ($0.05 \leq x \leq 1.00$) particle powders. As $x$ rises in the range $0.05 \leq x \leq 0.20$, there is an enhancement of the photoluminescence due to the establishment of an energy-looping. Increasing x further, there are clear excitation power thresholds from which the PA-like mechanism ensues. We demonstrate several upconversion lines, and found out that the degree of nonlinearity in the power dependences is related to $x$ and the photoluminescence channel. We estimate the spatial resolution that could be get for each $x$ and emission wavelength. Finally, we exploit the population redistribution among the $Nd^{3+}$ states, and demonstrate the operation of an optical thermometer with large relative thermal sensitivity.

## 2. EXPERIMENTAL

*a. Morphological and structural characterization*

The $Nd_xY_{1.00-x}Al_3(BO_3)_4$ particle powders, $0.05 \leq x \leq 1.00$, studied here are the same presented in Ref. [8], which were synthetized by the polymeric precursor method. Details are given in the Supplementary Material. The morphological and structural characterizations were previously performed: X-ray diffraction results revealed a rhombohedral structure with R32 space group (hexagonal cell) for powders of $0.05 \leq x \leq 0.80$, for materials with $x = 1.00$ the crystalline structure is monoclinic cell with C2/c space group [8]. The particle size distribution is broad, with its peak centered around 150 nm, and is almost independent of $x$ [8]. The maximum phonon energy is 1187 cm$^{-1}$ and almost not change as a function of $x$ [17,18], and the density of $Nd^{3+}$ for $x = 1.00$ is $5 \times 10^{21}$ ions/cm$^3$.

It should be noted that the $Nd_xY_{1.00-x}Al_3(BO_3)_4$ powders have been investigated in different contexts: in random lasers [7,12] including the demonstration of new optical effects [8,9,19]; and in photon avalanche, where only the stoichiometric powder ($x = 1.00$) was investigated [13,14]. Then, the aim of the present work was not studied before, that is investigate the influence of $Nd^{3+}$ content in the PA-like mechanism on $Nd_xY_{1.00-x}Al_3(BO_3)_4$ powders. Due to the richness of the excitation route at 1064 nm, the outcomes are not trivial as described below.

*b. Optical experiments*

A continuous-wave (cw) Nd:YAG laser at 1064 nm was used as the excitation source, with a Gaussian (TEM$_{00}$) spatial profile. The excitation power (P$_{exc}$) could be varied up to 3.7 W. The laser light was focused using a 10-cm biconvex lens directly onto the powder, which was gently placed in a metallic sample holder. The excited diameter at the powder surface was of ≈30 μm, much lower than the accommodated powder dimensions (a cylinder with 3 mm of radius and 3 mm in depth). The generated light was collected using a pair of lenses, and focused onto a multimode optical fiber coupled to a spectrometer equipped with a charge-coupled device (CCD) that allows simultaneous spectral measurements from 330 nm to 1180 nm, with a resolution of 2.0 nm. The CCD integration time was fixed to allow consecutive spectrum acquisition every 50 ms. A short-pass optical filter was used to reject the elastically scattered photons from the excitation laser. The interaction of the 1064 nm laser with Nd$^{3+}$ ions involves nonradiative relaxations that increase the particles temperature. This temperature was monitored with a FLIR-E40 thermal camera (detection range: -20 ≤ T ≤ 650 °C, accuracy of 1°C, 30 frames per second) simultaneously with the spectral measurements. It allowed to associate each photoluminescence spectrum with the correct particles temperature. The visualization of the excited area during the experiments allowed us to select spatial points to evaluate the temporal evolution of the temperature. As before [14], we opted to analyze the maximum temperature at the excited area. The photoluminescence spectra and the particles temperature evolve at time scales on the order of 1 s or larger, which is much larger than the response time of the spectrometer and the thermal camera. Then, the acquisition system is fast enough to cope with the photo-induced heating and spectral changes. It is important emphasizing that the particles temperature increased intrinsically due to phonon emissions, i.e., no external heating source was used, such as a heating state. We limited the P$_{exc}$ in order to avoid the irreversible coalescence from the particles due to the large attained temperatures. In the P$_{exc}$ range we worked, the results were reproducible, as verified by chopping the excitation laser and getting similar results.

## 3. RESULTS AND DISCUSSION

The nonresonant excitation at 1064 nm of Nd$^{3+}$ in Nd$_x$Y$_{1.00-x}$Al$_3$(BO$_3$)$_4$ (0.05 ≤ $x$ ≤ 1.00) particle powders was performed at room temperature. The excitation dynamics is quite complex presenting strong dependence with P$_{exc}$, and the Nd$^{3+}$ content in the Nd$_x$Y$_{1.00-x}$Al$_3$(BO$_3$)$_4$ particles. An intrinsic temperature rise was inferred due to phonon emissions in the Nd$^{3+}$ relaxation pathways, which alters the time evolution of photoluminescence as a result of several temperature-dependent processes. The temperature increase can enhance the absorption of excitation photons, increase nonradiative relaxations, as well as thermal excitation from a given energy level to higher states. Some of the photoluminescence spectra are given in Fig. 1 for all investigated values of *x* as a function of particle exposure time to the excitation beam. In the results of Fig. 1, $P_{exc}$ was chosen close to 2.000 W, but spectra for the entire range of $P_{exc}$ and for all *x* are shown in the Supplementary Material (Figure S3). For low $Nd^{3+}$ concentration (0.05 ≤ *x* ≤ 0.20, Fig. 1 a, b, and c), the emission intensities increase with time and then saturate on a time scale of 1 s. On the other hand, the time evolution is slower for *x* = 0.20, which is explained later. The photoluminescence increase with time is first due to the temperature influence on the phonon-assisted excitation of the $Nd^{3+}$, i.e., the absorption of photons increases the particles temperature, which in turn increases the photon absorption (*absorption of excitation photons → temperature increase with phonons generation → absorption of excitation photons*). Three emission bands are seen (890 nm, 810 nm, and 750 nm), but low intensity emissions are present at 600 nm, 660 nm, and 690 nm. In the Supplementary Material, we provide color maps for the spectral evolution in which the intensity is given in logarithmic scale to facilitate the visualization of the less intense emissions. The excitation pathways are described later. Similar results are observed for all investigated $P_{exc}$, i.e., the monotonic growth of the emissions and the particles temperature after exposure to the excitation laser (Fig. S3 in the Supplementary Material). On the other hand, for 0.40 ≤ x ≤ 1.00 (Fig. 1 d-g) the scenario changes: the time evolution slows down for x = 0.40 and then quicken for larger *x*. For x ≥ 0.40, when increasing $P_{exc}$, the photoluminescence dynamics is fast for low $P_{exc}$, slows down at a given threshold excitation power ($P_{th}$), and accelerates for higher $P_{exc}$ (Fig. S3 in the Supplementary Material). This behavior of the rise-time is characteristic of PA [20]. Intense emissions in visible wavelengths (660 nm, 600 nm, and 536 nm) are unveiled for *x* ≥ 0.40 and $P_{exc}$ ≥ $P_{th}$. In the temporal evolution, the particles temperature presents an inflection point (for *x* ≥ 0.40 and $P_{exc}$ ≥ $P_{th}$) indicating a change in the underlying mechanism of excitation [21], which is the PA-like phenomenon described below. Blue light (480 nm), green light (536 nm), and near-infrared emissions (1200 nm < λ < 2000 nm) were also detected (Fig. S2).

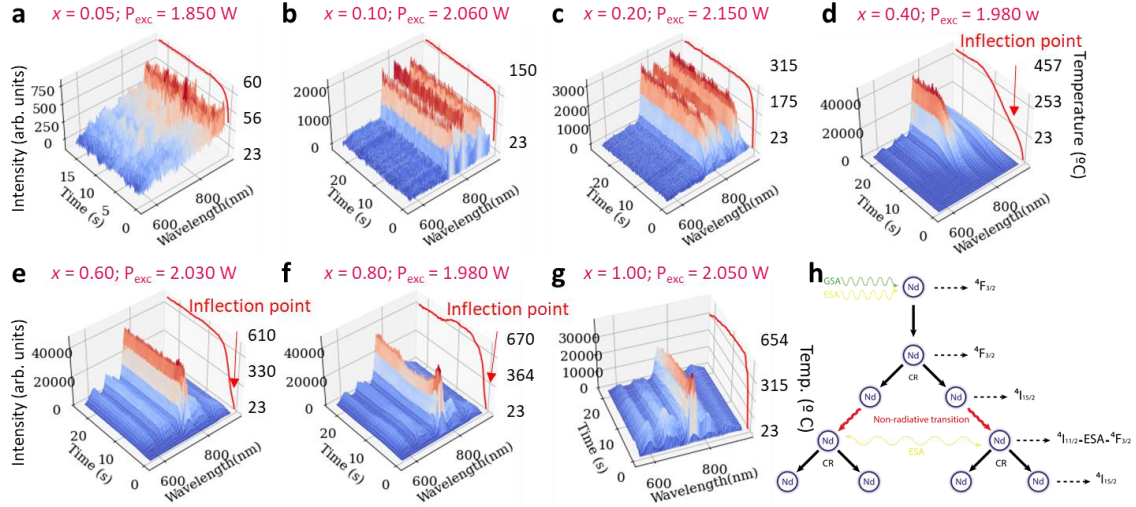

**FIG. 1.** Photoluminescence spectra and temperature of Nd$_x$Y$_{1.00-x}$Al$_3$(BO$_3$)$_4$ particles (0.05 ≤ x ≤ 1.00) are given as a function of the exposure time to the excitation laser at 1064 nm. The value of *x* as well as the excitation power (P$_{exc}$) are shown in the figure. Emissions bands are observed at 600 nm, 660 nm, 690 nm, 750 nm, 810 nm, and 890 nm associated with Nd$^{3+}$ 4f-4f elecronic transitions. Blue light at 480 nm, green light at 536 nm, and near-infrared emissions (1200 nm ≤ λ 2000 nm) were also measured[14], and the corresponding spectra are show in Fig. Sxxx. The particles temperature, which increased intrinsically due to light-to-heat conversion, is show in the right axis of each figure. (h) Illustration of the photon avalanche process, which starts with low probability absorption processes. After a sequence of cross-relaxation, nonradiative relaxations, and resonant excited-state absorption, the population of the Nd$^{3+}$ state $^4F_{3/2}$ is doubled every iteration.

The anti-Stokes excitation of Nd$^{3+}$ at 1064 nm is feasible with phonon annihilation from the Nd$_x$Y$_{1.00-x}$Al$_3$(BO$_3$)$_4$ lattice in order to compensate the energy mismatch with the ground-state transition $^4I_{9/2} \rightarrow {}^4F_{3/2}$. The excitation and emission pathways are shown in Fig. 2. Another excitation route involves the thermal excitation from the ground state to the first upper-lying level ($^4I_{11/2}$) followed by resonant excited-state absorption ($^4I_{11/2} \rightarrow {}^4F_{3/2}$) [22]. Both processes present low probability at room temperature. However, once the $^4F_{3/2}$ state is excited, the Nd$^{3+}$ can relax to lower states with photon emissions around 890 nm ($^4F_{3/2} \rightarrow {}^4I_{9/2}$), 1064 nm ($^4F_{3/2} \rightarrow {}^4I_{11/2}$), 1319 nm ($^4F_{3/2} \rightarrow {}^4I_{13/2}$), and 1800 nm ($^4F_{3/2} \rightarrow {}^4I_{15/2}$). The Nd$^{3+}$ at $^4F_{3/2}$ state can also relax nonradiatively by multiphonons to the $^4I_{15/2}$ level, whose rate is temperature-dependent. Another possibility to depopulate the emitter level $^4F_{3/2}$ is via cross-relaxation, in which the interaction with a neighboring ion at the ground state ($^4I_{9/2}$) occurs, ending with both ions at the $^4I_{15/2}$ state. This step is interesting because once at the $^4I_{15/2}$ level, fast phonon-assisted relaxations to the $^4I_{13/2}$ and, subsequently, to the $^4I_{11/2}$ are expected. Once at the $^4I_{11/2}$ state, the excitation beam is resonant with the ESA transition $^4I_{11/2} \rightarrow {}^4F_{3/2}$, promoting the two ions (initially at the $^4F_{3/2}$ and $^4I_{9/2}$ states) to the $^4F_{3/2}$ state. In resume, one ion excited at the $^4F_{3/2}$ by very weak excitation transitions can lead to two ions at this state after a cross-relaxation with a non-excited ion followed by nonradiative relaxations with phonon emissions and resonant ESA. That is, there is the establishment of an energy-looping in the sense that one excited ion can lead to two excited ions, the two ions can lead to four ions, and so on. Depending on the experimental conditions, this energy-looping can evolve to a PA (a representation is given in Fig. 1h), where stronger emissions with well-defined $P_{th}$ are expected [20]. Notice that nonradiative relaxations are associated with phonon emissions that increase the lattice temperature. This, in turn, enhances the GSA of excitation photons as well as the ESA due to the thermal excitation of the $^4I_{11/2}$ from the ground state.

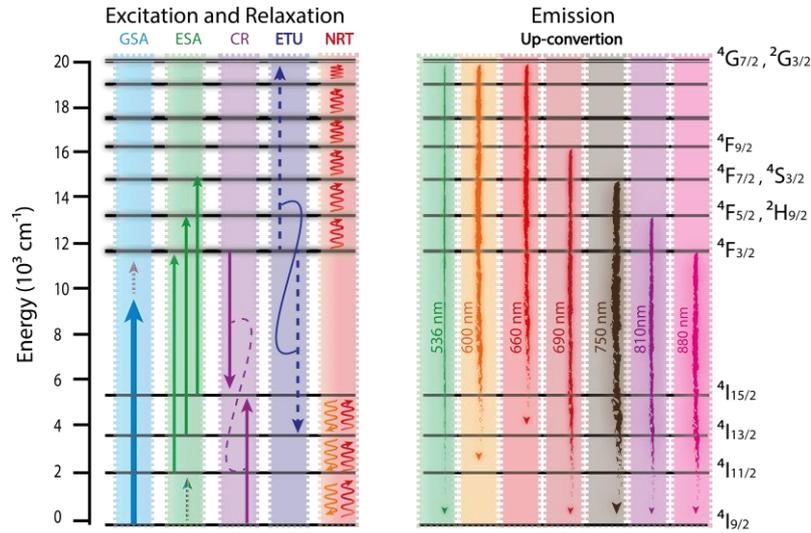

**FIG. 2.** Excitation (upward arrows) and emission (downward arrows) pathways of $Nd^{3+}$ at $Nd_xY_{1.00-x}Al_3(BO_3)_4$ particle powders. Thermal excitations and multiphonon decays between two given levels are represented by wavy arrows pointing up and down, respectively. GSA, ESA, CR, ETU, and NRT stand for Ground-State Absorption, Excited-State Absorption, Cross-Relaxation, Energy Transfer Upconversion, and Nonradiative Transitions, respectively.

Due to the cycle *absorption of excitation photons → temperature increase with the phonons generation → absorption of excitation photons*, the excitation dynamics evolve on a time scale larger (Fig. 1) than usually reported for $Nd^{3+}$ (tens of milliseconds) [23]. Because of that, the PA-like [13] process reported here is not a pure PA in the sense that the temperature increase plays a fundamental role to the energy amplification. The small slowdown of photoluminescence dynamics unveils the energy-looping among the $Nd^{3+}$ for $x \leq 0.20$, which triggers the PA-like behavior for $x \geq 0.40$ at well-defined threshold excitation powers ($P_{th}$).

The energy-looping as well as the PA, magnify the population of the $^4F_{3/2}$ state and, consequently, the emissions whose transitions originate from this level. However, losses of population are expected due to thermal excitation to upper-lying states [24–26], as well as energy transfer upconversion (ETU) between two ions at the $^4F_{3/2}$ [27,28]. The thermal excitation to the [$^4F_{5/2}$, $^2H_{9/2}$] generates light at around 810 nm ($^4F_{5/2}$, $^2H_{9/2}$ → $^4I_{9/2}$). The $Nd^{3+}$ at the [$^4F_{5/2}$, $^2H_{9/2}$] states can be excited thermally to the [$^4F_{7/2}$, $^4S_{3/2}$] whose ground-state relaxation results in emission at around 750 nm, and the thermal excitation [$^4F_{7/2}$, $^4S_{3/2}$] → $^4F_{9/2}$ can provide emission at around 690 nm ($^4F_{9/2}$ → $^4I_{9/2}$). This sequence of thermal excitations from the $^4F_{3/2}$ state to the high-lying is called ladder-thermal excitation, and it has been observed for Nd$^{3+}$ doped materials [24,29,30]. Another energy pathway for upconversion is the ETU between two Nd$^{3+}$ at the $^4F_{3/2}$ promoting one ion to the [$^4G_{7/2}$, $^2G_{3/2}$] states while the other one relaxes to the $^4I_{15/2}$ state. This latter relaxes nonradiatively to lower-lying states through the sequence $^4I_{15/2} \rightarrow {}^4I_{13/2} \rightarrow {}^4I_{11/2}$. Once at the $^4I_{11/2}$ state, the Nd$^{3+}$ can absorb one excitation photon and be promoted back to the $^4F_{3/2}$ state. On the other hand, the ions at the [$^4G_{7/2}$, $^2G_{3/2}$] states can relax to lower-lying levels radiatively as well as nonradiatively. Radiative relaxations to the $^4I_{9/2}$, $^4I_{11/2}$, and $^4I_{13/2}$ provides emissions at 536 nm, 600 nm, and 660 nm, respectively. In addition, the phonon-assisted relaxations can populate the $^4F_{9/2}$ state, from which relaxation to the ground-state generates light at 690 nm. The ions at the $^4F_{9/2}$ can also relax to the $^4F_{7/2}$ and $^4F_{5/2}$ contributing to the emissions at 810 nm and 880 nm, respectively. However, the populations of these states should be more influenced by thermal excitations from the $^4F_{3/2}$, since the corresponding emissions are observed for low $x$ and low $P_{exc}$ as well (Fig. 1, and Fig. S1 in the Supplementary Material). Notice the existence of two energy pathways to populate the [$^4G_{7/2}$, $^2G_{3/2}$] and $^4F_{9/2}$ states: ETU between two ions at the $^4F_{3/2}$ states and ladder-thermal excitations from the $^4F_{3/2}$ level. As discussed below, the former mechanism is dominant for low $P_{exc}$, while the latter prevails for large $P_{exc}$. There is also the possibility of excited state absorption from ions at the $^4F_{3/2}$ state, which can populate the [$^4G_{7/2}$, $^2G_{3/2}$] state. But, as previously demonstrated in bulk crystals of Nd$_x$Y$_{1.00-x}$Al$_3$(BO$_3$)$_4$, ETU is the dominant process over ESA [27,28]. Two-photon absorption from the excitation beam to populate the [$^4G_{7/2}$, $^2G_{3/2}$] states [31] should not be relevant here.

Examining the power dependence of the emissions (Fig. 3), one can infer about the underlying physical mechanisms for the population of different Nd$^{3+}$ energy levels. The data in log-log scale as well as the corresponding slopes are given in the Supplementary Material. For each $x$, we limited $P_{exc}$ to avoid coalescence of the particles due to the large attained temperature associated to the light to phonon conversion. Let us first focus our attention to samples with $0.05 \leq x \leq 0.20$. For $x = 0.05$ (Fig. 3a), the strongest emission is for 880 nm, whose transition departures from the first populated state ($^4F_{3/2} \rightarrow {}^4I_{9/2}$). Since the excitation of the $^4F_{3/2}$ state involves the annihilation of one excitation photon, the dependence of this 880 nm emission ($P_{880}$) with $P_{exc}$ is expected to be linear. However, it presents a nonlinear dependence (Fig. 3a) that is assigned to the energy-looping among the Nd$^{3+}$. Notice the similar dependences of $P_{880}$ and particles temperature with $P_{exc}$. Deviations from the linear behavior are also observed for emissions at 810 nm ($P_{810}$) and 750 nm ($P_{750}$), whose emitting states ($^4F_{5/2}$ and $^4F_{7/2}$, respectively) are populated via thermal excitation from the $^4F_{3/2}$ level. The relative intensities of the emissions with $P_{880} > P_{810} > P_{750}$ reinforce the ladder-thermal excitation $^4F_{3/2}$

→ $^4F_{5/2}$ → $^4F_{7/2}$. For *x* = 0.20, the dependences of the emissions $P_{880}$, $P_{810}$, and $P_{750}$ with $P_{exc}$ present larger nonlinearities in the set of samples with 0.05 ≤ *x* ≤ 0.20, which is assigned to the energy-looping among the $Nd^{3+}$. Under low $P_{exc}$, $P_{880}$ > $P_{810}$ > $P_{750}$, reverting under large $P_{exc}$, i.e., $P_{880}$ < $P_{810}$ < $P_{750}$. This is a consequence of the temperature increase with the $P_{exc}$, leading to population loss by thermal excitation from a given level to upper-lying states. At large $P_{exc}$ the leak of population from the $^4F_{3/2}$ state is so strong that $P_{880}$ presents some signs of saturation for *x* = 0.20. Notice also the emergence of intense visible emissions at 660 nm and 600 nm, whose emitting states ($^4G_{7/2}$, $^2G_{3/2}$) can be populated by thermal coupling with lower-lying levels and/or ETU between two $Nd^{3+}$ at the $^4F_{3/2}$ state. Since the emissions are only pronounced for large $P_{exc}$, which corresponds to high temperatures, one can wonder about an interplay of the excitation mechanism, i.e., at low $P_{exc}$ the ETU should be the dominant while for large $P_{exc}$ the ladder-thermal excitation ($^4F_{3/2}$ → $^4F_{5/2}$ → $^4F_{7/2}$ → $^4F_{9/2}$ → $^4G_{7/2}$, $^2G_{3/2}$) should overcome the ETU. Summarizing, for this first set of samples (0.05 ≤ *x* ≤ 0.20), the excitation laser populates the $^4F_{3/2}$ state by phonon assisted absorption transitions. More energetic emissions (shorter wavelengths) were observed with the interplay between ETU and ladder-thermal excitation from the $^4F_{3/2}$ state to upper-lying levels feasible by an intrinsic temperature increase due to phonon emissions in the $Nd^{3+}$ energy pathways. The corresponding emissions present nonlinear dependences with $P_{exc}$ due to an energy-looping, but there is no evidence of a threshold power from which the dependences change abruptly.

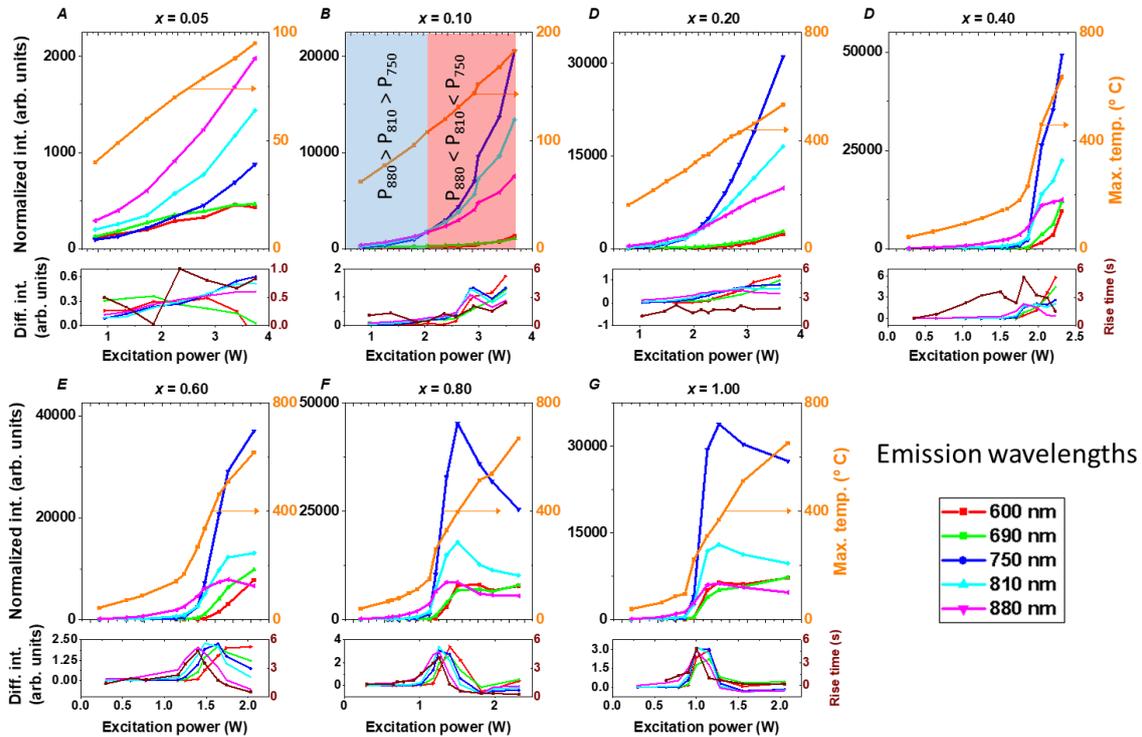

**FIG. 3.** Photoluminesncence power dependencies with the excitation power ($P_{exc}$) for diffenrent *x* in the $Nd_xY_{1.00-x}Al_3(BO_3)_4$ powders as well as the particles temperature, which increased intrinsically due to phonon emissions in the excitation parthway of the Nd$^{3+}$. The bottom part of each figure represente the derivatives of each emission line.

Moving to $0.40 \leq x \leq 1.00$, the power dependences show well-defined threshold excitation powers ($P_{th}$) from which the photoluminescence increase suddenly (Fig. 3). This is attributed to the PA mechanism, which is known to exhibit a $P_{th}$. The PA-like ensues for $x \geq 0.40$, and $P_{th}$ decreases with increasing $x$. For $P_{exc} < P_{th}$ ($0.40 \leq x \leq 1.00$), the output power relation is $P_{880} > P_{810} > P_{750}$, which reverses for $P_{exc} > P_{th}$. Notice also the appearance of intense visible emissions at large powers, as before ($0.05 \leq x \leq 0.20$), but much more intense here. In resume, the energy amplification depends on $x$: for $x \leq 0.20$ there is an energy-looping that precedes the PA-like observed for $x \geq 0.40$. Corroborating with the ladder-thermal excitation $^4F_{3/2} \rightarrow [^4F_{5/2}, ^2H_{9/2}] \rightarrow [^4F_{7/2}, ^4S_{3/2}] \rightarrow ^4F_{9/2}$, the saturation of emissions occurs progressively at high $P_{exc}$ as $x$ increases (880 nm $\rightarrow$ 810 nm $\rightarrow$ 750 nm $\rightarrow$ 690 nm). Also, there is an inversion in the relative intensities for $P_{600}$ and $P_{690}$: $P_{600}$ is stronger than $P_{690}$ for $x < 0.20$; at $x = 0.20$ both emissions present similar intensities; and for $0.40 \leq x \leq 0.60$ the relation reverses, i.e., $P_{690} > P_{600}$; finally, for $x > 0.60$ both emissions saturate with similar intensities. It allows us to conclude that increasing $x$ and for large $P_{exc}$, the population of the upper-lying states responsible for the more energetic emissions arises by ladder-thermal excitations from lower-lying levels, starting with the population of the $^4F_{3/2}$ state and becomes pronounced due to the large increase in the particles temperature. Finally, the emissions $P_{880}$, $P_{810}$, and $P_{750}$ decrease after a given $P_{exc}$ (Fig. 3, $0.40 \leq x \leq 1.00$), while the more energetics $P_{690}$ and $P_{600}$ saturate. The decrease of the emitted intensities at 880 nm, 810 nm, and 750 nm for $P_{exc} > P_{th}$ and $x \geq 0.60$ starts with loss of population from the corresponding emitting level to upper-lying states by thermal excitations. Furthermore, other important thermal excitations occur for lower-lying levels. First, the thermal excitation from the $^4I_{11/2}$ to the $^4I_{13/2}$ state. Amidst that, the resonant ESA $^4I_{11/2} \rightarrow ^4F_{3/2}$ rate decreases and, consequently, the emission at 880 nm. With the thermal population of the $^4I_{11/2}$ state, there is the possibility of almost resonant ESA from the $^4I_{13/2}$ to the $^4F_{5/2}$, which would increase the emission at 810 nm. While the temperature rises, the Nd$^{3+}$ at $^4I_{13/2}$ state are promoted to the $^4I_{15/2}$ level. Then, the emission at 810 nm decreases, and that one at 750 nm increases due to another almost resonant ESA transition $^4I_{15/2} \rightarrow [^4F_{7/2}, ^4S_{3/2}]$. This is supported by the continue increase of the particles´ temperature with $P_{exc}$ (Fig. 3).

In Fig. 1, we represented data for $P_{exc}$ close to 2.000 W because in the course of our measurements we did not attempt to take the same $P_{exc}$ from one powder to the other. Then, we represented in Fig. 1 excitation powers closed to 2.000 W. Only if the excitation power values were very close to $P_{th}$ in Fig. 3, one would expect some impact on the temperature and the photoluminescence spectra because the excitation regime could be different, i.e., above or below the PA-like threshold. However, there is no threshold ($x \leq 0.20$) or the applied $P_{exc}$ was above $P_{th}$ ($x \geq 0.40$). Moreover, during the data acquisition the $P_{exc}$ fluctuation was quite low and does not impact the analysis, i.e., there is not the possibility of $P_{exc}$ change randomly from above to below the $P_{th}$.

Given the multitude of effects in the excitation pathways of the $Nd^{3+}$, the $Nd_xY_{1.00-x}Al_3(BO_3)_4$ particles present complex dependences with $P_{exc}$. For low $x$ ($\leq 0.20$), there is the establishment of an energy-looping that increases the absorbance of the nonresonant excitation beam. By increasing $x$, in addition to the number of $Nd^{3+}$, which increases the photon absorbance by phonon-assisted transitions, the cross-relaxation rate between excited ($^4F_{3/2}$) and non-excited ($^4I_{9/2}$) levels also increases, which favors the energy-looping. That is due to the electric dipole nature of the interaction among $Nd^{3+}$ that increases with $r^{-6}$, where $r$ is the distance between the interacting ions. Also, nonradiative relaxations increase the particles temperature, which in turn favor the photons absorption as well as the thermal excitations from the $^4F_{3/2}$ level to upper-lying states resulting in emissions at several wavelengths. Contrary to several works that show the luminescence concentration quenching [1,8], the overall photoluminescence increases up to $x = 0.60$ thanks to the PA-like process, which is evident by the sudden increase of the photoluminescence at a given $P_{th}$. For $x > 0.60$, there is the triggering of the PA-like, but the emissions decrease or saturate at large $P_{exc}$. It is not a concentration quenching in the sensing that the proximity among the $Nd^{3+}$ favors energy migration, which is eventually lost to killer centers. But the rich energy level structure of the $Nd^{3+}$ close to the ground-state changes the excitation mechanism as the particles temperature increases intrinsically due to phonon emissions, and the complex interplay of ESA from the $Nd^{3+}$ lower-lying levels is unveiled only at large temperatures.

Aiming applications in image systems, we evaluated the relation $P_{xyz} \propto P_{exc}^S$, where $S$ stands for the number of annihilated photons to produce the emission at the wavelength $xyz$ nm. In order to do that, the data of Fig. 3 were represented in log-log scale (Fig. S2 from the Supplementary Material), and fitted with linear curves whose slopes give $S$. The values of $S$ as well as the predicted spatial resolution for each xyz nm and x ($\delta_{xyz,x}$) are given in Tables S1 and S2 in the Supplementary Material. The ratio between the predicted spatial resolution expected for one-photon excitation ($\delta_{xyz,x}/\delta_{one-photon}$) and under PA-like excitation at wavelength $xyz$ nm and $Nd^{3+}$ content ($x$) is given in Table S3. Resolutions greater than four times the diffraction limit are predicted for visible emissions at 600 nm. Due to the threshold and saturation behaviors, one can wonder about the useful range of $P_{exc}$. To evaluate that, the derivatives of the emissions are given in Fig. 3. Satisfactory tradeoff between large derivatives and the wide range $P_{exc}$ are get for $x = 0.40$ and $0.60$.

Applications in optical thermometry due to population redistribution is prominent. In the first report of PA-like mechanism, da Silva et al. [13] demonstrated that the ratio between the emissions at 810 nm and 880 nm presents large variation with the particles temperature, and relative thermal sensitivity of 2.95 %°C$^{-1}$ at 34 °C was obtained. Lately, Szalkowski et al. [32] performed an in-depth investigation on the impact of temperature on PA behavior in Tm$^{3+}$:NaYF$_4$ nanoparticles and 40 %K$^{-1}$ above 400 K ($\approx$ 15 %K$^{-1}$ at room temperature) was predicted. Here, we evaluated the luminescence intensity ratio for the emissions at 750 nm and 880 nm – see the Supplementary Material. The values of the relative thermal sensitivities (> 4.0 %°C$^{-1}$) are among the record values [33], but we anticipate that more studies will be necessary since the thermal/electronic response from the particles depend on P$_{exc}$.

It is worth mentioning that we are only pointing prominent applications for the PA-like mechanism. The main focus of the present article is to unveil the underlying excitation mechanisms, which populate several energy levels of the Nd$^{3+}$. Real-life applications require addressing, for example, how to dissociate the effects of environmental temperature and P$_{exc}$ on the PA-like as well as the photoluminescence. Recently, we could distinguish between intrinsic heating associate with Nd$^{3+}$ nonradiative relaxations and the temperature increase by an external heating source [14]. We showed the possibility of trigger the PA-like by heating the particles, while keeping P$_{exc}$ below the threshold at room-temperature. In theranostic applications, several fundamental aspects should be investigated, such as, the necessity of homogeneous particle sizes at the nanoscale, dispersibility in aqueous media, and potentially surface functionalization.

## 4. SUMMARY AND CONCLUSIONS

The complex dependences from the Nd$_x$Y$_{1.00-x}$Al$_3$(BO$_3$)$_4$ powders with the Nd$^{3+}$ content and P$_{exc}$, make the set of samples suitable for imaging systems, optical thermometers, and theranostics due to the possibility of excitation and emission channels fitting the biological windows as well as the simultaneous light-induced heating [34]. Largest nonlinearities (*S*) were obtained for *x* = 0.40 and *x* = 0.60 at the 600 nm and 690 nm emission wavelengths. The range of P$_{exc}$ within the nonlinear region depends on *x*. Then, the Nd$^{3+}$ content can be chosen to get a tradeoff between the largest *S* and the wide range of P$_{exc}$. What is more, the coexistence of several phenomena provides a multitude of emission wavelengths, which is advantageous because the emission channel can be chosen according to the detection system. Furthermore, optical thermometry is prominent due to the population redistribution among the Nd$^{3+}$ energy levels. Large relative thermal sensitivity with value larger than 4.0 %°C$^{-1}$ was obtained. Notice the biocompatibility of the excitation at 1064 nm [23]. Subsequently, the sensitivity of the photon-avalanche to environmental changes around the particles will be investigated soon aiming the development of ultra-sensitive sensors and, for this, optimal materials and excitation routes need to be found. Finally, the $Nd_xY_{1.00-x}Al_3(BO_3)_4$ particles allow the observation of second-harmonic generation, which can be exploited for optical thermometry with large sensitivity, as recently reported [35–37].

**ACKNOWLEDGEMENTS**

We acknowledge financial support from the Brazilian Agencies: Fundação de Amparo à Pesquisa do Estado de Alagoas (FAPEAL), Coordenação de Aperfeiçoamento de Pessoal de Nivel Superior (CAPES) – Finance Code 001, Fundação de Amparo à Pesquisa do Estado de Goiás (FAPEG), Financiadora de Estudos e Projetos (FINEP), Conselho Nacional de Desenvolvimento Científico e Tecnológico (CNPq) through the grant: Nr. 427606/2016-0; scholarships in Research Productivity 2 under the Nr. 303990/2019-8 and 305277/2017-0; National Institute of Photonics (INCT de Fotônica). C.V.T.M and R.F.S thank CNPq for their Scientific Initiation scholarship. We also thank to J. G. B. Cavalcante, M. A. Nascimento, and J. F. da Silva for technical support.

**SUPPLEMENTARY MATERIAL**

The spectra for each excitation power are presented as 3D as well as colormap plots. The data in Fig. 3 are presented in log-log scale. The dependence of $P_{th}$ with *x* is presented. Based on the experimental results, we calculated the spatial resolution for each $x \geq 0.20$. Aiming applications in optical thermometry, we determined the relative thermal sensitivity considering the emissions at 750 nm and 880 nm from the $Nd^{3+}$.

**CREDIT AUTHORSHIP CONTRIBUTION STATEMENT**

**R. F. da Silva:** Methodology, Data curation, Writing – original draft, and Editing. **D. F. Luz:** Methodology, Data curation, Writing – original draft, and Editing. **C. V. T. Maciel:** Methodology, Data curation, Writing – original draft, and Editing. **E. P. Santos:** Methodology, Data curation, Writing – original draft, and Editing. **G. Soares:** Methodology, Data curation, Writing – original draft, and Editing. **L. J. Q. Maia:** Particles synthesis, Methodology, Writing – original draft, and Editing. **C. Jacinto:** Methodology, Writing – original draft, and Editing. **A. L. Moura:** Methodology, Data curation, Writing – original draft, and Editing.

**DECLARATION OF COMPETING INTEREST**

The authors declare that they have no known competing financial interests or personal relationships that could have appeared to influence the work reported in this paper.

**DATA AVAILABILITY**

The data that support the findings of this study are available from the corresponding author upon reasonable request.

**REFERENCES**

1. A. Benayas, D. Jaque, J. G. Solé, N. I. Leonyuk, E. Bovero, E. Cavalli, and M. Bettinelli, "Effects of neodymium incorporation on the structural and luminescence properties of the YAl$_3$(BO$_3$)$_4$–NdAl$_3$(BO$_3$)$_4$ system," J. Phys. Condens. Matter **19**(24), 246204 (2007).
2. M. Ju, G. Sun, X. Kuang, C. Lu, Y. Zhu, and Y. Yeung, "Theoretical investigation of the electronic structure and luminescence properties for Nd$_x$Y$_{1-x}$Al$_3$(BO$_3$)$_4$ nonlinear laser crystal," J Mater Chem C **5**(29), 7174–7181 (2017).
3. S. T. Jung, J. T. Yoon, and S. J. Chung, "Phase transition of neodymium yttrium aluminum borate with composition," Mater. Res. Bull. **31**(8), 1021–1027 (1996).
4. E. G. Hilário, R. S. Pugina, M. L. da Silva-Neto, L. J. Q. Maia, J. M. A. Caiut, and A. S. Gomes, "Structural and morphological characterization of Y$_{1-x}$Nd$_x$Al$_3$(BO$_3$)$_4$ micron-sized crystals powders obtained by the urea precipitation method and its random laser properties," J. Lumin. **243**, 118624 (2022).
5. C. Jacinto, T. Catunda, D. Jaque, and J. G. Solé, "Fluorescence quantum efficiency and Auger upconversion losses of the stoichiometric laser crystal NdAl$_3$(BO$_3$)$_4$," Phys Rev B **72**(23), 235111 (2005).
6. S. Garcia-Revilla, I. Iparraguirre, C. Cascales, J. Azkargorta, R. Balda, M. A. Illarramendi, M. Ai-Saleh, and J. Fernandez, "Random laser performance of Nd$_x$Y$_{1-x}$Al$_3$(BO$_3$)$_4$ laser crystal powders," Opt Mat **34**(2), 461–464 (2011).
7. E. G. Rocha, Í. R. Paz, B. J. Santos, W. C. Soares, E. de Lima, L. M. Leão, L. J. Maia, and A. L. Moura, "Self-induced optical parametric amplification of random laser emission," Laser Phys **29**(4), 045402 (2019).
8. S. J. M. Carreño, A. L. Moura, Z. V. Fabris, L. J. Q. Maia, A. S. L. Gomes, and C. B. de Araújo, "Interplay between random laser performance and self-frequency conversions in Nd$_x$Y$_{1.00-x}$Al$_3$(BO$_3$)$_4$ nanocrystals powders," Opt Mat **54**, 262–268 (2016).
9. A. L. Moura, V. Jerez, L. J. Q. Maia, A. S. L. Gomes, and C. B. de Araújo, "Multi-wavelength emission through self-induced second-order wave-mixing processes from a Nd$^{3+}$ doped crystalline powder random laser," Sci Rep **5**, 13816 (2015).
10. A. L. Moura, L. J. Maia, V. Jerez, A. S. Gomes, and C. B. de Araújo, "Random laser in Nd:YBO$_3$ nanocrystalline powders presenting luminescence concentration quenching," J Lumin **214**, 116543 (2019).
11. J. G. Câmara, D. M. da Silva, L. R. Kassab, C. B. de Araújo, and A. S. Gomes, "Random laser emission from neodymium doped zinc tellurite glass-powder presenting luminescence concentration quenching," J Lumin 117936 (2021).
12. A. L. Moura, P. I. Pincheira, L. J. Maia, A. S. Gomes, and C. B. de Araújo, "Two-color random laser based on a Nd$^{3+}$ doped crystalline powder," J Lumin **181**, 44–48 (2017).
13. J. F. da Silva, R. F. da Silva, E. P. Santos, L. J. Maia, and A. L. Moura, "Photon-avalanche-like upconversion in NdAl$_3$(BO$_3$)$_4$ nanoparticles excited at 1064 nm," Appl Phys Lett **117**(15), 151102 (2020).
14. E. P. Santos, C. V. Maciel, R. F. da Silva, D. F. Luz, J. F. Silva, C. Jacinto, L. J. Maia, F. A. Rego-Filho, and A. L. Moura, "Temperature triggering a photon-avalanche-like mechanism in NdAl$_3$(BO$_3$)$_4$ particles under excitation at 1064 nm," J. Lumin. 118645 (2022).
15. A. Bednarkiewicz, E. M. Chan, A. Kotulska, L. Marciniak, and K. Prorok, "Photon avalanche in lanthanide doped nanoparticles for biomedical applications: super-resolution imaging," Nanoscale Horiz. **4**(4), 881–889 (2019).
16. C. Lee, E. Z. Xu, Y. Liu, A. Teitelboim, K. Yao, A. Fernandez-Bravo, A. M. Kotulska, S. H. Nam, Y. D. Suh, and A. Bednarkiewicz, "Giant nonlinear optical responses from photon-avalanching nanoparticles," Nature **589**(7841), 230–235 (2021).
17. H. Kim and S. Shi, "Raman spectra of monoclinic neodymium-aluminum borate," Chin. Phys. Lett. **5**(1), 1 (1988).
18. J. O. Pimenta, Z. V. Fabris, and L. J. Q. Maia, "Blue photoluminescence behavior in Tm$_x$Y$_{1-x}$Al$_3$ (BO$_3$)$_4$ nanopowders and structural correlations," Mater. Sci. Eng. B **247**, 114383 (2019).
19. A. L. Moura, S. J. M. Carreño, P. I. R. Pincheira, Z. V. Fabris, L. J. Q. Maia, A. S. L. Gomes, and C. B. de Araújo, "Tunable ultraviolet and blue light generation from Nd:YAB random laser bolstered by second-order nonlinear processes," Sci Rep **6**, 27107 (2016).
20. M.-F. Joubert, "Photon avalanche upconversion in rare earth laser materials," Opt. Mater. **11**(2–3), 181–203 (1999).
21. Y. Dwivedi, A. Bahadur, and S. B. Rai, "Optical avalanche in Ho:Yb:Gd$_2$O$_3$ nanocrystals," J. Appl. Phys. **110**(4), 043103 (2011).
22. K. Trejgis, K. Maciejewska, A. Bednarkiewicz, and L. Marciniak, "Near-Infrared-to-Near-Infrared Excited-State Absorption in LaPO4:Nd$^{3+}$ Nanoparticles for Luminescent Nanothermometry," ACS Appl. Nano Mater. **3**(5), 4818–4825 (2020).
23. L. Marciniak, A. Bednarkiewicz, and K. Elzbieciak, "NIR–NIR photon avalanche based luminescent thermometry with Nd$^{3+}$ doped nanoparticles," J. Mater. Chem. C **6**(28), 7568–7575 (2018).
24. L. de S. Menezes, G. S. Maciel, C. B. de Araújo, and Y. Messaddeq, "Thermally enhanced frequency upconversion in Nd$^{3+}$-doped fluoroindate glass," J Appl Phys **90**(9), 4498–4501 (2001).
25. A. Bednarkiewicz, D. Hreniak, P. Dereń, and W. Strek, "Hot emission in Nd$^{3+}$/Yb$^{3+}$: YAG nanocrystalline ceramics," J. Lumin. **102**, 438–444 (2003).
26. W. Strek, L. Marciniak, A. Bednarkiewicz, A. Lukowiak, D. Hreniak, and R. Wiglusz, "The effect of pumping power on fluorescence behavior of LiNdP$_4$O$_{12}$ nanocrystals," Opt. Mater. **33**(7), 1097–1101 (2011).
27. D. Jaque, J. Capmany, F. Molero, Z. D. Luo, and J. G. Solé, "Up-conversion luminescence in the Nd$^{3+}$: YAB self frequency doubling laser crystal," Opt. Mater. **10**(3), 211–217 (1998).
28. D. Jaque, O. Enguita, Z. D. Luo, and J. G. Solé, "Up-conversion luminescence in the NdAl$_3$(BO$_3$)$_4$(NAB) microchip laser crystal," Opt. Mater. **25**(1), 9–15 (2004).
29. M. S. Marques, L. de S. Menezes, W. Lozano B, L. R. P. Kassab, and C. B. De Araujo, "Giant enhancement of phonon-assisted one-photon excited frequency upconversion in a Nd$^{3+}$-doped tellurite glass," J Appl Phys **113**(5), 053102 (2013).
30. K. Kumar, D. K. Rai, and S. B. Rai, "Observation of ESA, ET and thermally enhanced frequency upconversion in Nd$^{3+}$:LiTeO$_2$ glass," Eur. Phys. J.-Appl. Phys. **41**(2), 143–149 (2008).
31. V. K. Rai, C. B. de Araújo, Y. Ledemi, B. Bureau, M. Poulain, and Y. Messaddeq, "Optical spectroscopy and upconversion luminescence in Nd$^{3+}$ doped Ga$_{10}$Ge$_{25}$S$_{65}$ glass," J Appl Phys **106**(10), 103512 (2009).
32. M. Szalkowski, M. Dudek, Z. Korczak, C. Lee, L. Marciniak, E. M. Chan, P. J. Schuck, and A. Bednarkiewicz, "Predicting the impact of temperature dependent multi-phonon relaxation processes on the photon avalanche behavior in Tm$^{3+}$: NaYF$_4$ nanoparticles," Opt. Mater. X **12**, 100102 (2021).

33. J. Stefańska, A. Bednarkiewicz, and L. Marciniak, "Advancements in excited state absorption-based luminescence thermometry," J. Mater. Chem. C (2022).
34. G. Chen, H. Qiu, P. N. Prasad, and X. Chen, "Upconversion Nanoparticles: Design, Nanochemistry, and Applications in Theranostics," Chem. Rev. **114**(10), 5161–5214 (2014).
35. J. F. da Silva, C. Jacinto, and A. L. Moura, "Giant sensitivity of an optical nanothermometer based on parametric and non-parametric emissions from $Tm^{3+}$ doped $NaNbO_3$ nanocrystals," J. Lumin. 117475 (2020).
36. T. Zheng, M. Runowski, I. R. Martín, S. Lis, M. Vega, and J. Llanos, "Nonlinear Optical Thermometry—A Novel Temperature Sensing Strategy via Second Harmonic Generation (SHG) and Upconversion Luminescence in $BaTiO_3$: $Ho^{3+}$, $Yb^{3+}$ Perovskite," Adv. Opt. Mater. 2100386 (2021).
37. T. Zheng, M. Runowski, P. Woźny, B. Barszcz, S. Lis, M. Vega, J. Llanos, K. Soler-Carracedo, and I. R. Martín, "Boltzmann vs. non-Boltzmann (non-linear) thermometry-$Yb^{3+}$-$Er^{3+}$ activated dual-mode thermometer and phase transition sensor via second harmonic generation," J. Alloys Compd. **906**, 164329 (2022).